\documentclass[runningheads]{llncs}
\usepackage{amsmath,amssymb,amsfonts}
\usepackage{algorithmic}
\usepackage{svg}
\usepackage{graphicx}
\usepackage{textcomp}
\usepackage{multirow}
\usepackage{makecell}
\usepackage{xcolor}
\usepackage{caption}
\usepackage{tabularx}
\usepackage[ruled,linesnumbered]{algorithm2e}
\usepackage{enumitem}
\usepackage{caption}

\title{Iterative Explainability for Weakly Supervised Segmentation in Medical PE Detection}

\author{Florin Condrea*\inst{1,2}\and
Saikiran Rapaka\inst{3} \and
Marius Leordeanu\inst{1,2,4}}
\authorrunning{Condrea Florin et al.}
\titlerunning{iExplain: Iterative Explainability for PE Segmentation}
\institute{Institude of Mathematics of the 
Romanian Academy ”Simion Stoilow",
Bucharest, Romania \and
Foundational Technologies, Siemens,
 Brasov, Romania \and  Siemens Healthineers,
 Princeton, USA \and  Polytechnic University of Bucharest,
 Bucharest, Romania
\email{florin.condrea@siemens.com}}

\usepackage{listings}

\definecolor{changecolor}{RGB}{0,0,0}
\definecolor{changecolor2}{RGB}{0,0,0}

\begin{document}

\maketitle
\begin{abstract}

Pulmonary Embolism (PE) are a leading cause of cardiovascular death. Computed tomographic pulmonary angiography (CTPA) is the gold standard for PE diagnosis, with growing interest in AI-based diagnostic assistance. However, these algorithms are limited by scarce fine-grained annotations of thromboembolic burden. We address this challenge with iExplain, a weakly supervised learning algorithm that transforms coarse image-level annotations into detailed pixel-level PE masks through iterative model explainability. Our approach generates soft segmentation maps used to mask detected regions, enabling the process to repeat and discover additional embolisms that would be missed in a single pass. This iterative refinement effectively captures complete PE regions and detects multiple distinct embolisms. Models trained on these automatically generated annotations achieve excellent PE detection performance, with significant improvements at each iteration. We demonstrate iExplain's effectiveness on the RSPECT augmented dataset, achieving results comparable to strongly supervised methods while outperforming existing weakly supervised methods.

\end{abstract}

\section{Introduction}

Pulmonary embolism (PE), characterized by the deposition of blood clots in the pulmonary arteries, are a critical public health challenge, ranking as the third most prevalent cardiovascular syndrome globally\cite{raskob2014thrombosis}. PE incidence ranges from 39-115 per 100,000 individuals, contributing to approximately 300,000 annual US deaths\cite{keller2020trends,wendelboe2016global}. This health burden is increasing, partly due to associations with Covid-19 infections\cite{Katsoularis2022} and a documented upward trend in PE occurrence\cite{keller2020trends,lehnert2018acute,de2014trends}. The urgency of PE diagnosis is underscored by high early mortality rates, with a significant percentage of deaths occurring before treatment can be administered\cite{cohen2007venous}. This, combined with increasing hospital workloads\cite{kocher2011workload1,portoghese2014workload2}, demands rapid and accurate patient triage and prioritization systems.

CT pulmonary angiography (CTPA) is the diagnostic gold standard for PE detection\cite{oldham2011ctpa}, positioning this challenge within the scope of computer vision applications. Deep neural networks (DNNs), particularly convolutional neural networks (CNNs)\cite{lecun1998pattern,krizhevsky2017imagenet,ronneberger2015unet}, have proven to be powerful tools for visual pattern recognition. Their effectiveness is demonstrated across various medical imaging applications, including diagnosing conditions like  Covid-19\cite{polsinelli2020covid19}, and intracranial hemorrhage using CT scans\cite{prevedello2017automated}, as well as in other modalities such as radiography\cite{soffer2019radiology1,yamashita2018radiology2} and magnetic resonance imaging (MRI)\cite{ali2019mri2,lin2018mri1}.

\begin{figure}[t]
\includegraphics[width=1.0\textwidth]{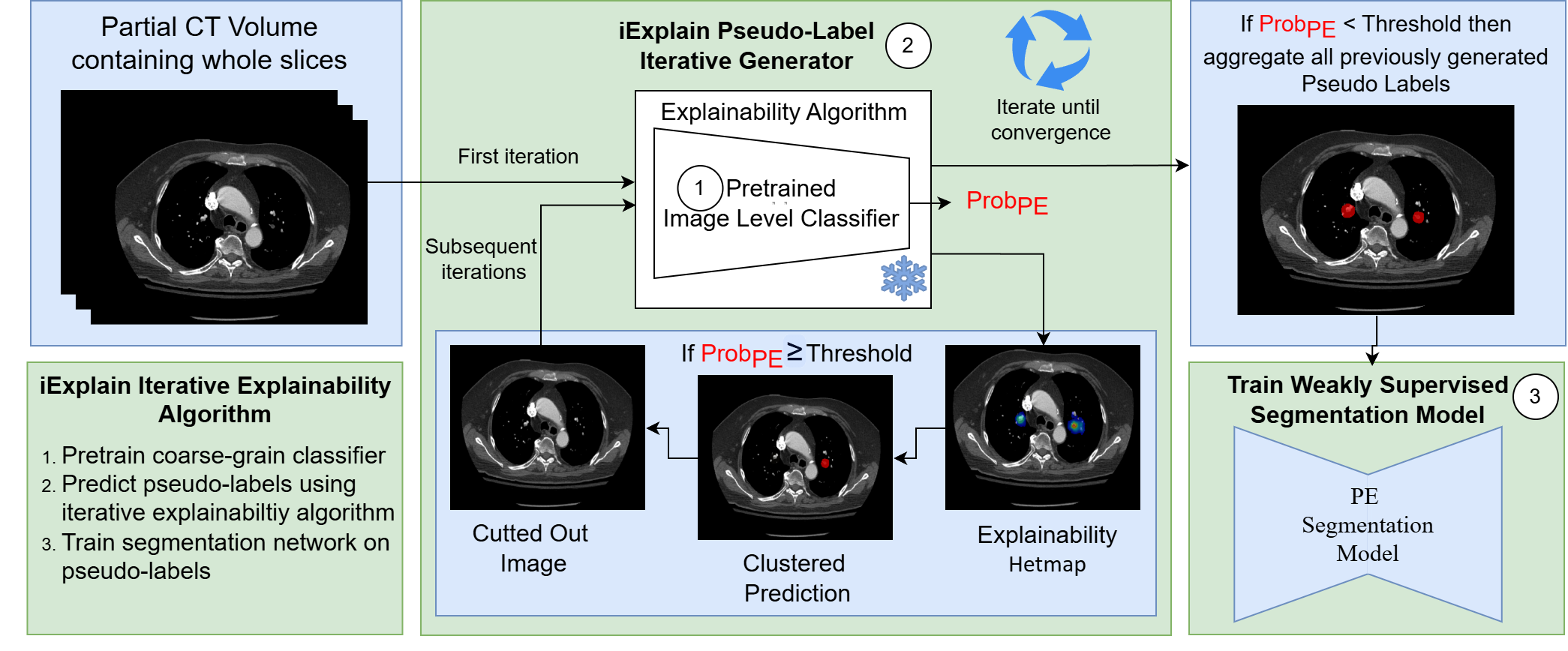}

\caption{iExplain, our weakly supervised algorithm for PE detection. 3D whole image mini-volumes (7 slices) are classified; positive cases generate PE probability heatmaps via Integrated Gradients. These are converted to clusters using Hysteresis thresholding, clusters are masked out in the input, and the process iterates until negative classification. Complete study predictions are obtained via sliding window and union of all detected clusters.}
\label{fig:IterativeExplainability}
\end{figure}

While the topic of segmentation in medical tasks has been well covered \cite{nnUnet2019,ronneberger2015unet,milletari2016vnet} and many recent publications cover the topic of PE classification either on slice or study level \cite{Cheikh2022AIDOC2,Soffer2021,Weikert2020AICOD1,Huang2020,Ma2022RSNAMultitask,condrea2024anatomically}, development of detailed localization or segmentation of the PEs \cite{ozkan2014novel_loc26,Weikert2020AICOD1,pu2023automated_peloc_wsl} is lagging behind. This could be attributed to the scarcity of fine-grained annotations for this task due to the expensive annotation process. The recently released public RSPECT augmented large scale dataset \cite{callejas2023augmentation_rspect} for PE localization will allow more contribution in this space.

\textbf{Main Contributions:} Our primary contributions are:
\begin{itemize}[noitemsep]
    \item \textbf{Novel Iterative Pseudo-Label Generation Algorithm:} We introduce a weakly supervised algorithm called \textbf{iExplain}, which leverages iterative model explainability for generating fine-grained PE segmentation pseudo-labels from image-level annotations. Its iterative refinement process progressively fully uncovers multiple PEs and crucially, operates without auxiliary models.
    \item \textbf{Strong PE Detection performance:} Our method yields strong PE detection and localization results on the RSPECT augmented dataset, demonstrating its dual utility:
    \begin{itemize}[itemsep=0pt] 
        \vspace{-1px}
        \item As a \textit{Weakly Supervised Learning (WSL) framework}, our model, trained solely on generated pseudo-labels, achieves a 71.6\% F1 score, matching a contemporary strongly supervised method \cite{zhu2024automatic_peloc3}.
        \vspace{-1px}
        \item As a \textit{strong pretraining strategy}, fine-tuning our WSL model with minimal human-annotated data boosts its F1 score to 75.5\%, significantly outperforming training from random initialization with the same limited data and highlighting a practical path to high performance.
    \end{itemize}
\end{itemize}

\section{Related Work }

Previous research on PE Computer Assisted Detection (CAD) systems follows two main approaches: early studies employed traditional image processing with segmentation and thresholding methods\cite{Zhou2009,Bouma2009,Pichon2004}, while recent work leverages Deep Learning, particularly CNNs\cite{Cheikh2022AIDOC2,Soffer2021,Huang2020,Weikert2020AICOD1}. Current research focuses on PE localization\cite{ozkan2014novel_loc26,zhu2024automatic_peloc3,xu2023automatic_peloc2,Weikert2020AICOD1}, enabling more fine-grained performance evaluation.

Model Explainability has become a point of interest in machine learning in recent years, many approaches \cite{sundararajan2017integrated,selvaraju2017gradcam,smilkov2017smoothgrad} being developed to best capture the sources of information in an input. While widely adopted for qualitative examples, recent work \cite{gonzalez2020iterative,fruh2021wsl_example,yang2022weakly_example2,nguyen2019novel_wsleye_e2e} have adopted feature use heatmaps generated by explainability methods and refined them into weakly supervised segmentations. Recent weakly supervised segmentation methods \cite{dang2022vessel,li2022weakly,xu2025weakly} have demonstrated success across different contexts and domains.

There are two papers of special interest for our work \cite{gonzalez2020iterative,pu2023automated_peloc_wsl}. Gonzalez et al. \cite{gonzalez2020iterative}, for instance, explored an iterative weakly supervised approach for 2D retinal disease segmentation, where explainability outputs were refined and an auxiliary model was employed for inpainting detected regions.

In the latter, Pu et al. \cite{pu2023automated_peloc_wsl} also train a deep learning model on pseudo-labels for PE localization, offering a comparison point. Their pseudo-label generation, however, is fundamentally distinct, heavily relying on anatomical priors and an external pulmonary artery segmentation model, within which classical computer vision algorithms identify PE candidates. In contrast, our proposed algorithm relies on an iterative process based on model explainability to generate our PE pseudo-labels.

\begin{figure*}[t]
\centering
\includegraphics[width=1.0\textwidth, height=0.62\textwidth]{ 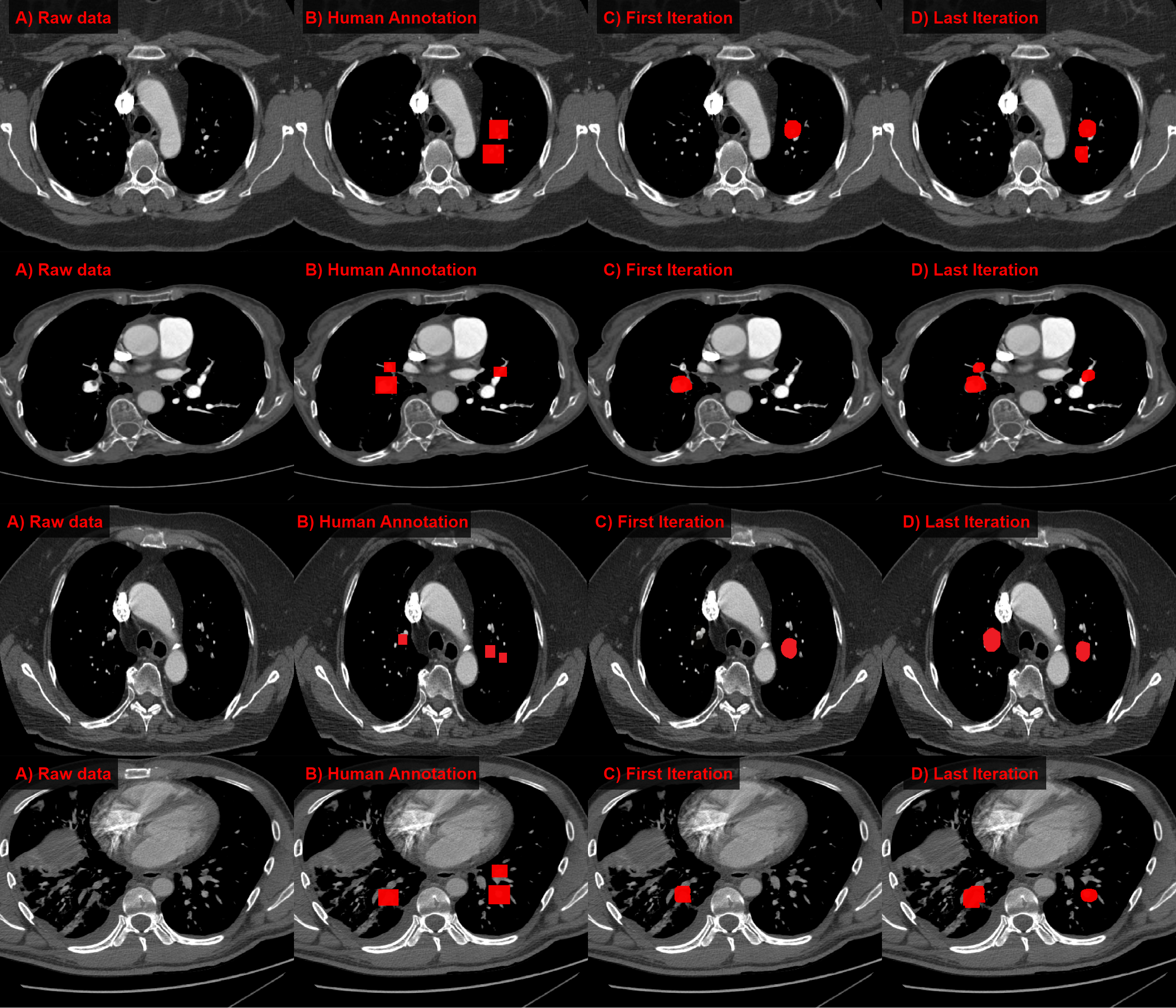}
\caption{Iterative refinement examples showing iExplain's PE discovery progression: A) Original study B) Ground Truth C) First iteration D) Final iteration. Our method detects both proximal PEs (rows 1-2) and PEs in the opposite lung (rows 3-4), demonstrating effective discovery of multiple embolisms through progressive masking.}
\label{fig:IterativeCase}
\end{figure*}

\section{Our method}

Our iterative weakly supervised learning algorithm, iExplain, follows an offline teacher - online student paradigm to train segmentation models from image-level annotations. The approach, illustrated in Fig. \ref{fig:IterativeExplainability}, consists of three stages to be presented in detail in the following section:
\vspace{-5px}
\begin{enumerate}[noitemsep]
\item Classification model for image level classification, trained on human annotations.
\item Iterative explainability module for weakly supervised segmentation generation, acting as an offline teacher, to obtain fine-grained weakly supervised annotations.
\item Segmentation model for pixel level classification, acting as an online student, trained on fine-grained WSL annotations.
\end{enumerate}

\textbf{Implementation details.} Our method is implemented in Python 3.8 and PyTorch 1.5.0, with significant reliance on the MONAI 1.1.0 \cite{cardoso2022monai} and Captum 0.6 \cite{kokhlikyan2020captum} libraries. Model training and inference were performed on an NVIDIA A100 node equipped with 4 GPUs.

\subsection{Slice level Classifier.} 
As suggested in recent publications \cite{condrea2024anatomically,Xu2021RSNA1st,than2020Kaggle3}, a 2.5D CNN has been employed as an image level classifier. We closely follow the baseline training recipe from a recent solution \cite{condrea2024anatomically} trained and tested on the RSPECT PE classification dataset \cite{RsnaPEDataset}, where a 2.5D EfficientNetV2-L is trained.  The input of the network is a 3D minivolume containing consecutive slices, resulting in a shape of (512,512,7), and the target label corresponding is the label of the center slice. Of special interest from the baseline \cite{condrea2024anatomically} for improving explainability performance of our algorithm is the augmentation scheme and final pooling layer used. The augmentation scheme contains an aggressive form of cutout augmentations \cite{devries2017cutout} taking out random parts of the input, which they show improves performance significantly, while also allowing input masking during our second stage. The pooling layers is MaxPool, which as noted by previous work \cite{solovyev20223maxpool}, improves the granularity of explainability methods, allowing for more focused explainability heatmaps.

\subsection{iExplain: Iterative Explainability pseudo-labels generation.} 

We generate weakly supervised labels using Integrated Gradients (IG) \cite{sundararajan2017integrated}, chosen for its effectiveness with small objects like PEs as indicated by prior work \cite{gonzalez2020iterative}. For each 3D mini-volume, IG and our slice-level model produce a 3D heatmap where high-intensity regions, corresponding to the PE itself \cite{wittram2007_pesign}, indicate influential areas for PE-positive predictions.

To convert these heatmaps into segmentations, we apply a Hysteresis-like thresholding method. This uses high and low thresholds: pixels above the high threshold are included, and pixels above the low threshold are included if connected to high-threshold pixels within a defined neighborhood (15x15x5), chosen based on geometric PE characteristics. This approach, while potentially missing parts of thin long tailed PEs, is robust for clustered PEs. The high threshold is optimized on the evalaution set, with the lower threshold set as half of it.

Finally, to create a heatmap for the entire CTPA volume, this process is applied in a sliding window centered on each slice. Predictions are aggregated as a Gaussian-weighted ($\sigma=0.8$) sum of neighboring slices.

\textbf{Iterative refinement} To improve the predictions, especially in the case of multiple PEs in a mini-volume as in Fig. \ref{fig:IterativeCase}, we adopt iterative refinement. Multiple cluster predictions are performed, after each iteration, predicted clusters are masked out of the CTPA volume. This process is repeated until either no more PEs are detected, new segmentations have a volume under 50 voxels or 10 iterations have been reached. First condition would mean no more steps are required, and the other two conditions are required as computational restrictions. The final segmentation is obtained through the union of all the prediction over the iterations.

\textbf{Prediction voting} To further improve the performance of the explainability module, we employ two heatmap-level voting techniques by generating multiple predictions for each volume. A first form of augmentation leverages references, a secondary input taken by integrated gradients \cite{sundararajan2017integrated}. A second form of augmentation is SmoothGrad \cite{smilkov2017smoothgrad}, where the input image has Gaussian noise added, and the final prediction is averaged over multiple noise instances. Both mechanisms demonstrated to improve the robustness of integrated gradients by previous work \cite{sturmfels2020visualizing_distil}. Our iterative refinement loop is described in Algorithm \ref{alg:1}.

\begin{algorithm}[t]
\caption{ Our iExplain Iterative Explainability Segmentation Algorithm}
\label{alg:1}
  \SetAlgoLined
  \KwIn{vol, clf, expl\_mod, clf\_thresh, heatmap\_thresh, iter\_limit}
  \KwOut{agg\_seg}
  segs $\gets$ []\, curr\_vol $\gets$ vol\, iter $\gets$ 0\, masked $\gets$ $\infty$\;

  \While{(clf.predict(curr\_vol) $>$ clf\_thresh) \textbf{and} (iter $<$ iter\_limit) \textbf{and} (masked $>$ heatmap\_thresh)}{
    iter $\gets$ iter + 1\;
    
    heatmap $\gets$ expl\_mod.explain(clf, curr\_vol)\;
    
    seg $\gets$ clustering(heatmap)\;
    
    masked $\gets$ sum(seg)\;
    
    curr\_vol $\gets$ curr\_vol * (1 - seg)\;
    
    segs.append(seg)\;
  }
  
  agg\_seg $\gets$ union(segs)\;

\end{algorithm}

\textbf{Pseudo-label filtering.}To reduce false positives, we use a high-sensitivity pseudo-label algorithm operating point and then post-process clusters. We apply empirically determined size and location filters: clusters smaller than a size threshold (to remove sporadic predictions) or beyond a 2D Euclidean distance threshold from the study center (to limit search to the lung area) are removed.

\subsection{Deep learning segmentation model.} 
For our third stage, we use a DynUNET (Monai framework \cite{cardoso2022monai}, variant of nnUnet \cite{nnUnet2019}) as the segmentation model. To enhance robustness to pseudo-label noise, we employ Generalized Dice Loss \cite{sudre2017generalise_dice} and Mixup data augmentation \cite{zhang2017mixup}. The model processes (128,128,64) voxel patches (HU windowed: center 100, width 400) to output PE probability heatmaps. Full-study heatmaps are generated using a sliding window (step 64,64,32) with Gaussian-weighted ($\sigma=0.25$) prediction aggregation. Training utilized augmentations from Stage 1 (excluding Cutout, adding RandomCrop to (128,128,64)), balanced sampling, and Generalized Dice Loss. The model was trained for 50,000 iterations (batch size 32) using AdamW optimizer \cite{loshchilov2017decoupled_adamw} with a learning rate of 1e-4, warmup, and linear decay to 1e-7.

\begin{table}[t]
  \centering
  \caption{Slice predictor performance. Our first stage model achieves AUCROC performance in the state-of-the-art range, indicating it is adequately trained for the following stage.}
   \label{tab:Classifier}
    \vspace{7px}
  \begin{tabular}{lccc}
    \textbf{Model} & \textbf{Dataset size} & \textbf{F1} & \textbf{AUCROC} \\
    Xu et al. \cite{Xu2021RSNA1st} & 6,279 & - & 96.4 \\
    Ajmera et al. \cite{ajmera2022deep_peslice2} & 942 & 37.0 & 94.0 \\
    Ma et al. \cite{Ma2022RSNAMultitask} & 5,292 & - & 93.0 \\
    \textbf{Ours} & 6,824 & \textbf{75.1} & \textbf{97.1} \\
  \end{tabular}
\end{table}

\subsection{Metric}

For evaluating our PE slice classification in the first algorithm stage, we employ AUC-ROC and F1 score, widely recognized metrics that comprehensively capture model performance. In the 2nd and 3rd stages focused on PE localization via segmentation, we utilize sensitivity, PPV, and F1 metrics at the finding/PE level, consistent with metrics used in leading research \cite{Weikert2020AICOD1,xu2023automatic_peloc2,zhu2024automatic_peloc3,pu2023automated_peloc_wsl}. We implement the cluster matching approach from Pu et al. \cite{pu2023automated_peloc_wsl}, whose pseudo-label training protocol closely resembles ours and serves as our primary comparison point alongside our ablation studies. This matching considers a detection and PE matched when they intersect, with matched pairs counted as true positives (TPs) and unmatched annotations or detections as false negatives (FNs) or false positives (FPs), respectively. Zhu et al.\cite{zhu2024automatic_peloc3} also employ this metric.

Due to RSPECT augmented dataset's \cite{callejas2023augmentation_rspect} use of bounding box annotations rather than precise PE outlines, segmentation metrics (Dice, IOU) are inappropriate. While mathematically equivalent to Dice, we use F1 for localization to avoid confusion with segmentation metrics. Fig. \ref{fig:Oversegmentation_overlayed} shows the mismatch between bounding boxes and actual PE segmentations.

\section{Results}
We train all our 3 stages on the RSPECT dataset \cite{RsnaPEDataset}, adding an additional private dataset to stage 3, and evaluate the performance on the RSPECT augmented dataset \cite{callejas2023augmentation_rspect}. 

\begin{figure}[b]
\centering
\includegraphics[width=0.48\textwidth]{ 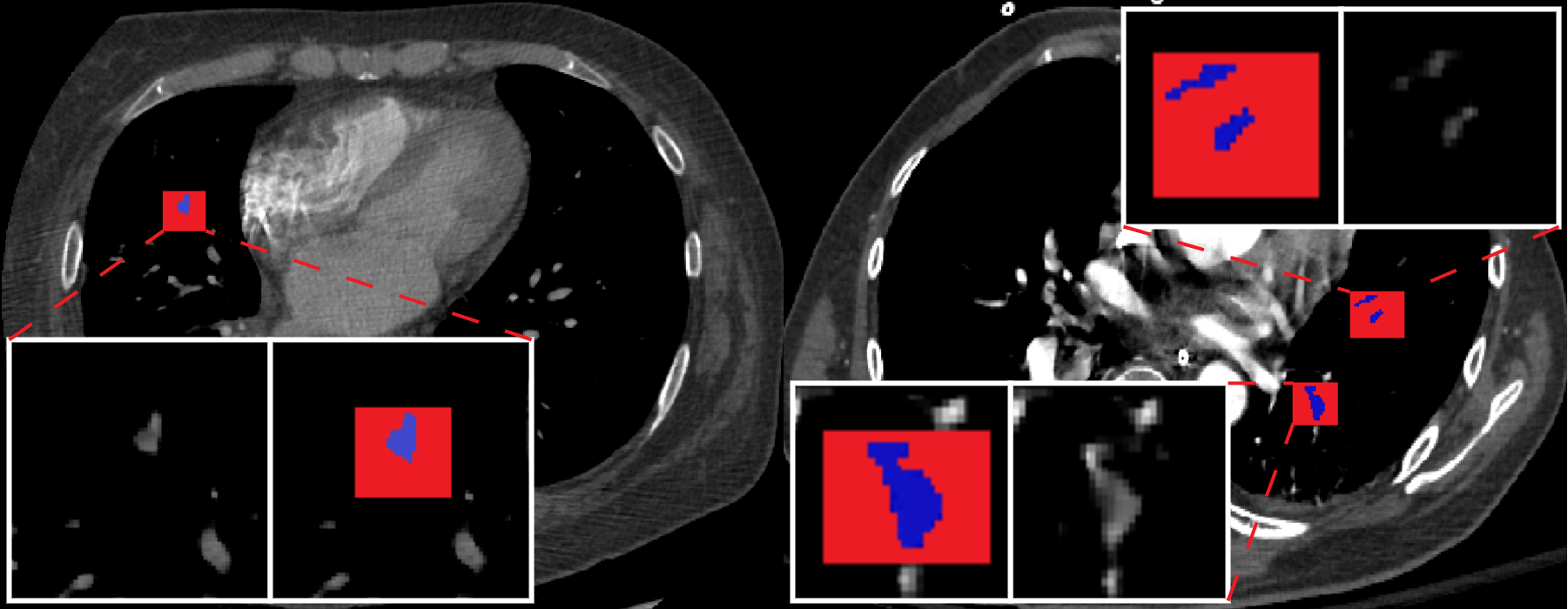}
\caption{Mismatch example between red bounding box annotations and blue PE segmentations. While bounding boxes work well for object detection evaluation, they're inadequate for segmentation evaluation. Therefore, we use F1 score for bounding box matching rather than segmentation Dice.}
\label{fig:Oversegmentation_overlayed}
\end{figure}

\textbf{Datasets.} RSPECT dataset \cite{RsnaPEDataset} has been released by the Radiological Society of North America as part of a Kaggle challenge \cite{rsna-str-pulmonary-embolism-detection}, contains 12,195 studies (CT scan volumes) with a positivity rate of 30.4\%, with slice-level annotations regarding the presence of a pulmonary embolism, and several other study-level annotations. A subset of 7279 studies have been published with annotations as part of a Kaggle challenge, further referred to as RSPECT challenge training dataset. The RSPECT \cite{callejas2023augmentation_rspect} augmented released subsequently contains a subset of 445 PE positive studies from the initial RSPECT challenge training dataset \cite{RsnaPEDataset}. For each study, PEs are annotated in each slice using a 2D bounding box, and a total of 30243 bounding boxes have been annotated, which we converted to segmentation masks despite some over-segmentation issues (Fig. \ref{fig:Oversegmentation_overlayed}).

For our experiments, we trained our models on RSPECT challenge training and our private data. To avoid data bleeding, studies also present in RSPECT augmented have been excluded from training, resulting in a training set of 6824 studies, with a positivity of 25.8\%. We used the RSPECT augmented dataset for testing both classification and localization, as it contains higher quality, finer-grained annotations.  We augmented our training set with an additional private dataset containing 6,005 studies, of which 27\% were positive cases.

\begin{figure}[]
\centering
\includegraphics[width=0.40\textwidth, height=0.32\textwidth]{ 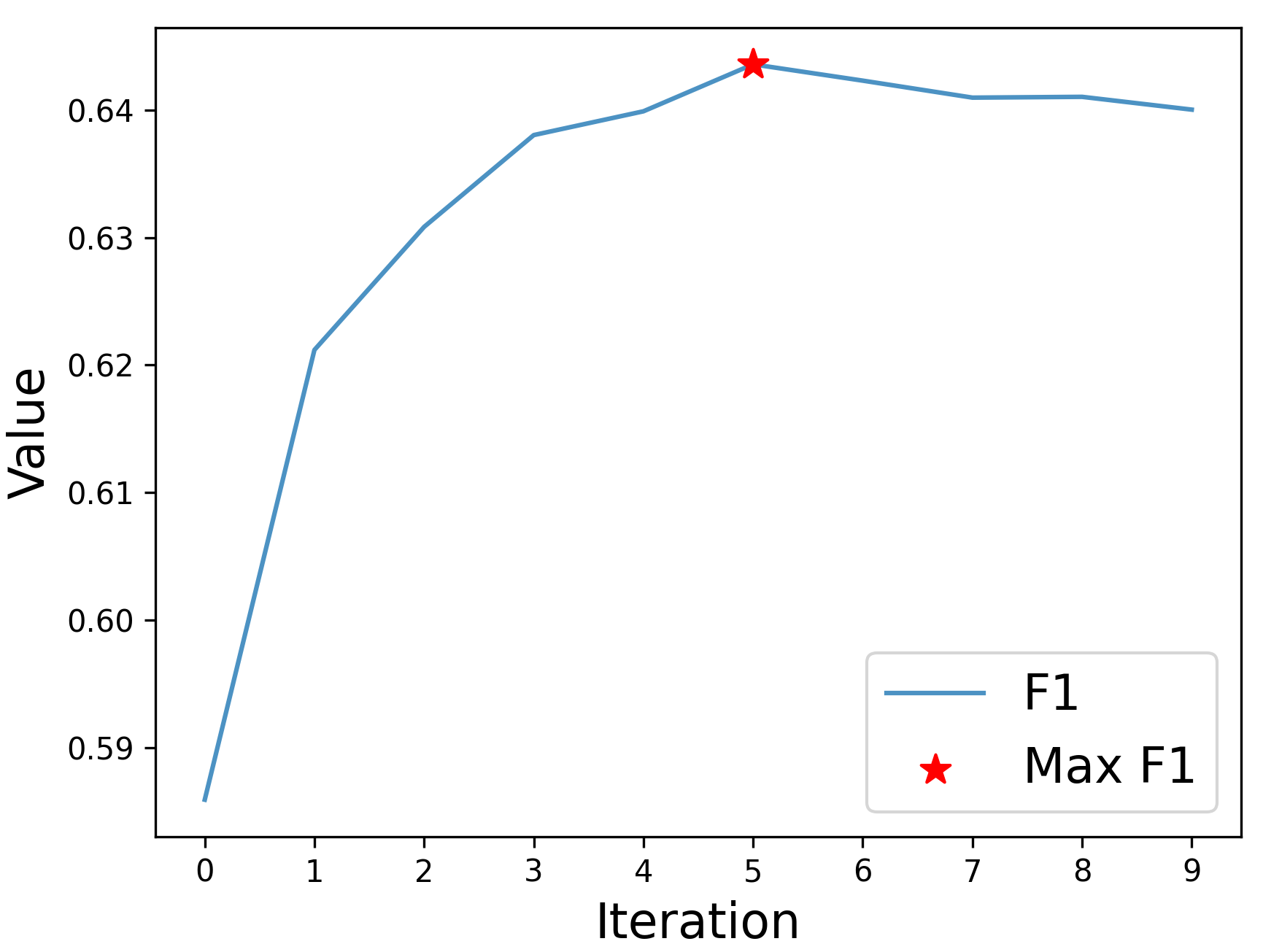}
\includegraphics[width=0.40\textwidth, height=0.32\textwidth]{ 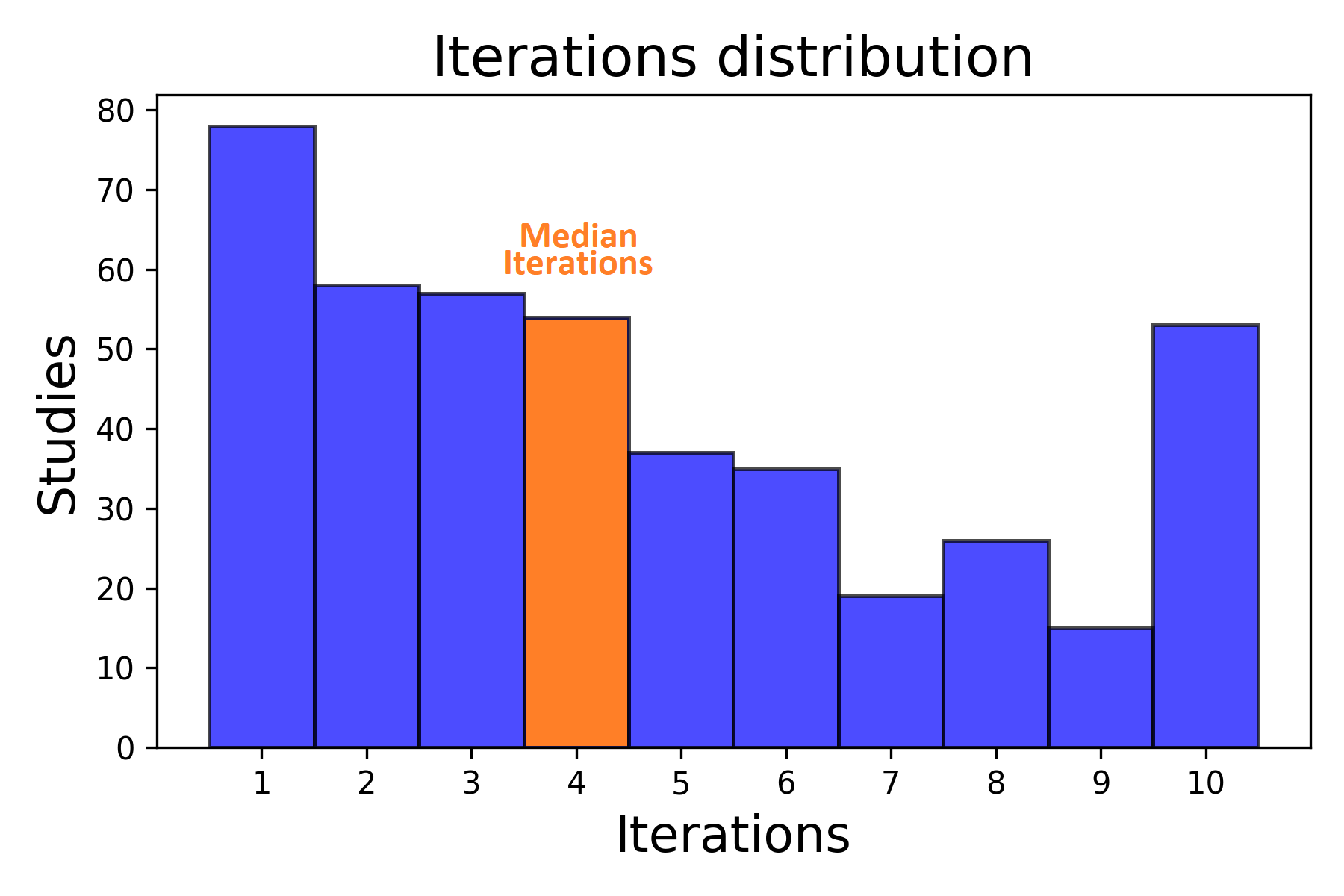}
\vspace{-3px}
\caption{\textbf{Left:} F1 performance evolution through iterations of iExplain. Sensitivity increases while PPV decreases as new PEs are discovered, reaching optimal F1 after 6 iterations before slight decline as metrics diverge. \textbf{Right:} Distribution of required refinement iterations in iExplain. Most studies need few iterations (~15\% need none), with decreasing requirements as PE detection completes. The 10-iteration limit creates a tail peak of cases that would continue refining despite diminishing performance.}
\label{fig:f1_recall_precision_new}
\end{figure}

\subsection{Image level classifier.} 
For the initial stage, we trained our 2.5D classifier on 6,824 studies containing slice level PE annotations, obtaining a well performing model for the next stage. Due to testing only on the fully positive RSPECT augmented dataset \cite{callejas2023augmentation_rspect}, marked as validation set, we report only slice-level metrics, excluding study level metrics. We obtain a slice-level performance AUCROC 97.08\% and an F1 score of 75.11\%, while on different test datasets, similar to other state-of-the-art methods \cite{Xu2021RSNA1st}. Further comparisons shown in Table \ref{tab:Classifier}.

\subsection{Weakly supervised segmentation.} 
We report PE localization performance on our validation set, RSPECT augmented \cite{callejas2023augmentation_rspect}, via bounding box matching for our weakly supervised model, with ablation studies showing improvements from each component.

\begin{table}[b]
    \caption{Stage 2 iterative refinement in iExplain significantly improves performance. Masking and reiterating discovers new PEs, increasing sensitivity while reducing PPV. F1 score peaks when these metrics balance.}
    \label{tab:wsl_iterative}
    \vspace{5px}
    \centering
    \begin{tabular}{ccccc}
    \textbf{References} & \textbf{Iteration}                       & \textbf{Sensitivity} & \textbf{PPV}  & \textbf{F1}     \\ 
    1          & \multicolumn{1}{c}{1}     &	50.8 &	68.8 & 58.4 \\
    1          & \multicolumn{1}{c}{10, best}  & 63.5 &	64.2 & 63.8 \\ \hline
    5          & \multicolumn{1}{c}{1}     &	48.3 &	74.5 & 58.6\\
    5          & \multicolumn{1}{c}{6, best}  & 58.4 &	71.6 & \textbf{64.4}

    \end{tabular}
    
\end{table}

\begin{table}[]
    \caption{Stage 2 cluster filtering. iExplain algorithm initially operates at high sensitivity, balanced by subsequent cluster filtering. This filtering significantly improves F1 scores across both single and multi-reference Integrated Gradients approaches.}
    \label{tab:wsl_filtering}
    \vspace{1px}
    \centering
    \begin{tabular}{ccccc}
        \textbf{References} & \textbf{Filtered}  & \textbf{Sensitivity} & \textbf{PPV} & \textbf{F1} \\
        1 & $\times$  & 65.1 & 45.1 & 53.3 \\
        1 & $\checkmark$  & 63.5 & 64.2 & 63.9 \\\hline
        5 & $\times$  & 59.3 & 49.0 & 53.6 \\
        \textbf{5} & \textbf{$\checkmark$}  & \textbf{58.4} & \textbf{71.6} & \textbf{64.4}  \\
    \end{tabular}
\end{table}

\textbf{Iterative refinement.} Table \ref{tab:wsl_iterative} demonstrates that iterative refinement in iExplain yields significant F1 score improvements of 5.8\%, consistent across both 1-reference and 5-reference Integrated Gradients scenarios. These gains stem from discovering new PEs by masking previous detections, evidenced by substantial sensitivity increases (6.7\% and 10.1\% respectively). While additional clusters introduce more false positives (reducing PPV by 4.6\% and 2.9\%), the sensitivity gains provide greater overall benefit. As shown in Fig. \ref{fig:f1_recall_precision_new}, F1 score peaks after 6 refinement iterations when sensitivity and PPV are balanced, before declining as these metrics diverge. The most substantial performance improvements occur in the initial iterations, with most cases requiring only a few refinement steps. Our adaptive stopping criterion terminates the process when no new PEs are detected, with approximately 15\% of cases needing no refinement. We applied a max limit of 10 iterations due to diminishing returns, visible in the histogram's final bar showing cases that would otherwise require additional processing. An example of additional PE discovery via refinement is also displayed in Fig. \ref{fig:IterativeCase}.

\textbf{Prediction filtering.} Table \ref{tab:wsl_filtering} demonstrates that postprocessing predicted clusters through size and location filtering substantially improves F1 score by 10.8\%. We maximize this improvement by initially setting a high-sensitivity operating point to capture candidate PEs for subsequent filtering, removing clusters smaller than 100 voxels (eliminating tiny predictions) and those beyond 160 pixels from the center (restricting detection to the lung area).

\begin{table}[t]
\vspace{-15px}
\caption{Deep learning model performance. Our iExplain model is trained in two data regimes of 6,824 and 13,329 CTPA studies, and evaluated on RSPECT augmented \cite{callejas2023augmentation_rspect} 445 studies. We slightly outperform comparable pseudo-label methods\cite{pu2023automated_peloc_wsl} on similar data and, with private data augmentation, match the performance of strongly supervised methods\cite{zhu2024automatic_peloc3} using identical evaluation criteria. \textcolor{blue}{Weikert et al \cite{Weikert2020AICOD1}}, the only strongly supervised method performing better than us is trained in a much richer data regime. * WSL methods.}
\label{tab:comparison_seg}
\vspace{5px}
\centering
\small
\begin{tabular}{cccc|cccc}
\textbf{Model}                            & \textbf{Data} & \textbf{Positives} & \textbf{F1}   & \textbf{Model}                                    & \textbf{Data} & \textbf{Positives} & \textbf{F1}   \\
\multicolumn{4}{c|}{\textbf{Weakly Supervised Methods}}                                        & \multicolumn{4}{c}{\textbf{Strongly Supervised Methods}}                                               \\
Pu et al \cite{pu2023automated_peloc_wsl} & 6,415         & 1990               & 69.1          & Ozkan et al \cite{ozkan2014novel_loc26}           & 142           & 142                & 67.7          \\
\multirow{2}{*}{Our Model}                & 6,824         & 1,766              & 69.4          & Tajbakhsh et al\cite{tajbakhsh2019computer_loc41} & 121           & 121                & 49.4          \\
                                          & 13,329        & 3,389              & \textbf{71.6} & Xu et al \cite{xu2023automatic_peloc2}            & 113           & 113                & 66.1          \\
\textbf{}                                 &               &                    &               & Zhu et al \cite{zhu2024automatic_peloc3}          & 142           & 142                & 71.6          \\
                                          &               &                    &               & Our Model                                         & 13,329        & 3,389              & \textbf{71.6} \\
                                          &               &                    &               & Weikert et al \cite{Weikert2020AICOD1}            & 30,000        & 15,858             & \textbf{85.8} \\
                                          &               &                    &               &                                                   &               &                    &              
\end{tabular}
\end{table}

\subsection{Deep learning segmentation model.} 
For our final stage, we present model localization performance, results displayed in Table \ref{tab:comparison_seg}. Our main comparison is another PE algorithm \cite{pu2023automated_peloc_wsl} that, the same as us, train on generated PE pseudo-labels on the same data from RSPECT dataset \cite{RsnaPEDataset}, however their dataset has a higher positivity of 31.0\% than our 25.8\% due to us removing the RSPECT augmented \cite{callejas2023augmentation_rspect} from our training data. When trained on similar data volumes, we obtain an improvement of 0.3\% F1 score, obtaining F1 scores of 69.4\% compared to their 69.1\% . We also augment our train data using private data, and obtain 71.6\% F1 score, an increase of 2.2\%. Given the slight differences in dataset compositions and possible mismatches in evaluation implementations, the performances are not directly comparable.  

We also compare with fully supervised methods trained on human made annotations \cite{Weikert2020AICOD1,xu2023automatic_peloc2,zhu2024automatic_peloc3}. We directly compare with Zhu et al \cite{zhu2024automatic_peloc3} on the same matching metric. They also utilize a different subset of RSPECT \cite{RsnaPEDataset}, indicating a similar data distribution. We obtain the same F1 score performance of 71.6\%. While they train on much less data, although human annotated, they also rely on additional models to generate pulmonary artery segmentation, helping exclude common False Positive cases of PEs detected outside the arteries \cite{zhu2024automatic_peloc3}. While F1 scores are similar, their operating point leads to a  higher sensitivity of 86.0\% and lower PPV of 61.3\% compared to our model.

Xu et al\cite{xu2023automatic_peloc2} 
report using a stricter version of our matching criteria, requiring an IOU of over 50\%, and the matching criteria of Weikert et al\cite{Weikert2020AICOD1} is unknown.

\begin{figure}[b]
\centering
\includegraphics[width=0.96\textwidth,height=0.2\textwidth]{ 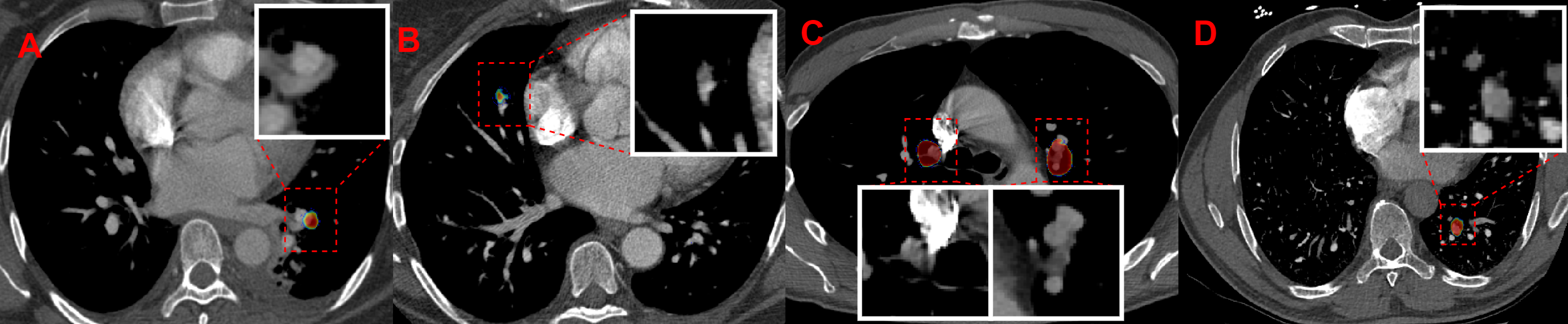}
\caption{Qualitative Error Examples. Common PE detection false positives include: A) Reduced contrast in a vein adjacent to an artery is mistaken for a PE. B) High contrast in veins due to poor contrast timing. C) Image Left: Shadows due to imaging artifact, Image Right: vein mistaken for an artery. D) Vein mistaken for an artery.}
\label{fig:qualitative_examples}
\end{figure}

\begin{table}[t]
\caption{Strongly supervised training results. Finetuning our iExplain pseudo-label trained model on human annotations yields a 3.8\% F1 score improvement, outperforming both random initialization (by 4.5\% F1) and Zhu et al.'s \cite{zhu2024automatic_peloc3} method trained on comparable data. Our baseline iExplain WSL model (italicized) shows the gains achieved through finetuning. Models trained/fine-tuned on 25\%, respectively tested on 75\% of RSPECT augmented \cite{callejas2023augmentation_rspect}.}
\label{tab:comparison_finetune}
\vspace{5px}
\centering
\small
\setlength{\tabcolsep}{4pt}
\centering
\begin{tabularx}{0.850\textwidth}{>{\raggedright\arraybackslash}c rrrr r}
\multicolumn{1}{c}{\textbf{Model}} & \textbf{Data} & \textbf{Positives} & \textbf{Recall} & \textbf{PPV} & \textbf{F1} \\
\textit{iExplain (ours) baseline} & \textit{13,329} & \textit{3,389} & \textit{68.3} & \textit{75.4} & \textit{71.7} \\ \hline 
Zhu et al \cite{zhu2024automatic_peloc3} & 142 & 142 & \textbf{86.0} & 61.3 & 71.6 \\
Our network Random Init & 111 & 111 & 66.2 & 76.6 & 71.0 \\
iExplain (ours) Finetunned & 111 & 111 & 73.9 & \textbf{77.5} & \textbf{75.5} \\
\end{tabularx}
\end{table}

\textbf{Strongly supervised finetune.} To evaluate our WSL iExplain algorithm as a pretraining strategy, we finetuned our model on 111 human-annotated studies (25\% of RSPECT Augmented) and evaluated on the remaining 334 studies in RSPECT augmented, comparable to Zhu et al.'s dataset. We also trained an identical model from scratch on the same data to isolate the contribution of pretraining.
Finetuning our WSL-pretrained model yielded a 3.8\% F1 score improvement, enhancing both sensitivity and PPV. This finetuned model outperformed Zhu et al.'s more complex algorithm by 3.9\% F1 despite using similar training data. Furthermore, our WSL-pretrained model surpassed the randomly initialized model obtaining F1 71.0\%, by 4.5\%, confirming our approach's effectiveness as a pretraining strategy. This enables developing strong models from coarse annotations that can be further refined with limited fine-grained human annotations. Table \ref{tab:comparison_finetune} details these results.

For finetuning, we used a shorter schedule of 10,000 iterations (batch size 32) with AdamW optimizer, learning rate 1e-5, 200-iteration warmup, and linear decay to 1e-7. Random initialization training used 25,000 iterations, learning rate 1e-4, 500-iteration warmup.

\textbf{Error Analysis} Like other PE detection methods\cite{Weikert2020AICOD1,zhu2024automatic_peloc3}, our model struggles with challenging cases. Common false positives include contrast issues, imaging artifacts, and confusion between pulmonary veins and arteries, while false negatives show no clear pattern. Error examples are displayed in Fig \ref{fig:qualitative_examples}.

%
\section{Conclusions}

We introduced a novel approach for weakly supervised segmentation based on an iterative procedure in which an explainability module provides a soft segmentation map, which is then used to mask the input for the next iteration, where the pseudo label generation using the explainability module is repeated. In this manner we can effectively capture the full extent of PE regions in the slice, as it is well known that such explainability modules do not detect the full object of interest. Moreover our method also allows for the detection of multiple distinct regions. We have extensively evaluated our method and showed that the performance improves significantly iteration by iteration, reaching a performance that is in the top two among all strongly supervised methods and significantly better than all published weakly supervised ones. While easy to implement and elegant, our approach can open new doors in weakly supervised and unsupervised segmentation research, with a wide application in the larger computer vision domain.

\textbf{Acknowledgements:} Marius Leordeanu was supported
in part by projects "Romanian Hub for Artificial Intelligence - HRIA", Smart Growth, Digitization and Financial Instruments Program, 2021-2027 (MySMIS No. 334906) and "European Lighthouse of AI for Sustainability - ELIAS", Horizon Europe program (Grant No. 101120237).

\textbf{Disclaimer:} The concepts and information presented in this paper/presentation are based on research results that are not commercially available. Future commercial availability cannot be guaranteed.

\bibliography{cas-refs}
\bibliographystyle{splncs03_unsrt}

\end{document}